\documentclass[superscriptaddress,nofootinbib,twocolumn,aps]{revtex4-2}

\usepackage{comment}
\usepackage{epsfig,amsmath,amssymb,amsfonts,color,algorithm,algcompatible,amsthm,bm,braket}
\usepackage{multirow}
\usepackage[table,xcdraw]{xcolor}
\usepackage{graphicx}
\usepackage{subfigure}
\usepackage{float}
\usepackage[normalem]{ulem}

\usepackage{algorithm}
\usepackage{algpseudocode}

\begin{document}

\preprint{APS/123-QED}

\title{Adaptive measurement strategy for noisy quantum amplitude estimation with variational quantum circuits}

\author{Kohei~Oshio}
\affiliation{Mizuho Research \& Technologies, Ltd., 2-3, Kandanishiki, Chiyoda-ku, Tokyo, 100-8233, Japan}
\affiliation{Quantum Computing Center, Keio University, 3-14-1 Hiyoshi, Kohoku-ku, Yokohama, Kanagawa, 223-8522, Japan}

\author{Yohichi~Suzuki}
\affiliation{Quantum Computing Center, Keio University, 3-14-1 Hiyoshi, Kohoku-ku, Yokohama, Kanagawa, 223-8522, Japan}
\affiliation{Global Research and Development Center for Business by Quantum-AI Technology (G-QuAT), National Institute of Advanced Industrial Science and Technology (AIST), 1-1-1, Umezono, Tsukuba-shi, Ibaraki 305-8568, Japan}

\author{Kaito~Wada}
\affiliation{Department of Applied Physics and Physico-Informatics, Keio University, Hiyoshi 3-14-1, Kohoku-ku, Yokohama 223-8522, Japan}

\author{Keigo~Hisanaga}
\affiliation{Department of Applied Physics and Physico-Informatics, Keio University, Hiyoshi 3-14-1, Kohoku-ku, Yokohama 223-8522, Japan}

\author{Shumpei~Uno}
\affiliation{Mizuho Research \& Technologies, Ltd., 2-3, Kandanishiki, Chiyoda-ku, Tokyo, 100-8233, Japan}

\author{Naoki~Yamamoto\thanks{
		e-mail address: \texttt{yamamoto@appi.keio.ac.jp}
	}}
\affiliation{Quantum Computing Center, Keio University, 3-14-1 Hiyoshi, Kohoku-ku, Yokohama, Kanagawa, 223-8522, Japan}
\affiliation{Department of Applied Physics and Physico-Informatics, Keio University, Hiyoshi 3-14-1, Kohoku-ku, Yokohama 223-8522, Japan}

             
\begin{abstract}
In quantum computation, amplitude estimation is a fundamental subroutine that is utilized in various quantum algorithms. 
A general important task of such estimation problems is to characterize the estimation lower bound, which is referred to as quantum Cramér–Rao bound (QCRB), and to construct an optimal estimator that achieves QCRB. 
This paper studies the amplitude estimation in the presence of depolarizing noise with unknown intensity. 
The main difficulty in this problem is that the optimal measurement depends on both the unknown quantum state and the amplitude we aim to estimate. 
To deal with these issues, we utilize the variational quantum circuits to approximate the (unknown) optimal measurement basis combined with the 2-step adaptive estimation strategy which was proposed in the quantum estimation theory. 
We numerically show that the proposed method can nearly attain the QCRB. 
\end{abstract}

\maketitle

\section{Introduction}
In quantum computation, amplitude estimation~\cite{Brassard2002} is a fundamental algorithm applied to diverse applications, such as Monte Carlo integration~\cite{Montanaro2015, Rebentrost2018, Woerner2019, Stamatopoulos2020, Miyamoto2020}, machine learning~\cite{Wiebe2016, Kerenidis2019}, and quantum chemistry~\cite{Knill2007, Wang2019, Lin2020, Wang2021, Dong2022}.  
Currently quantum computing devices lack tolerance against errors caused by external noises.
Therefore, several proposals incorporate noise models into the estimation process, including those that reduce the necessary resources (number of qubits and gate operations)~\cite{Y.Suzuki2020, Nakaji2020, Grinko2021, Tanaka2021, Herbert2021, Tiron2022, Tanaka2022}.
To analyze the precision of such estimators, 
the quantum estimation theory~\cite{Helstrom1976,Holevo2011} provides powerful tools; 
that is, for general estimation problems of unknown parameters embedded in quantum states (or quantum channels), quantum estimation theory can be used to discuss the fundamental estimation limits and estimation strategies. 
In particular, the ultimate lower bound of estimation error, independent of the measurement, is called the quantum Cram\'{e}r-Rao bound (QCRB)~\cite{Helstrom1968,Braunstein1994,Paris2009}.

For the noisy amplitude estimation problem, few studies investigate the achievability of QCRB~\cite{Uno2021, Wada2022}. 
However, these methods assume that the system is influenced by known depolarization noise, and they showed that the optimal bound is attainable in the limit of a large number of qubits. 
Clearly, the next research subject is to develop a noisy amplitude estimation algorithm that can handle unknown noise intensity parameters.

In this paper, we study the noisy amplitude estimation problem where the noise source is assumed to be depolarization, but its intensity is unknown; thus the problem is to estimate both the amplitude and the noise intensity. 
This problem contains two major difficulties. 
First, in general, the QCRB of the parameter of interest (the amplitude in our case) may become worse than the QCRB obtained when the nuisance parameters are known. 
The second difficulty is that the optimal measurement used to construct the optimal estimator is unknown. 
More specifically, the optimal measurement basis ($\ket{\lambda_0}$ and $\ket{\lambda_1}$ in Eq.~(\ref{EQ_optimal_measurement_basis})) contains unknown basis sets ($\ket{\psi_0}_n\ket{0}$ and $\ket{\psi_1}_n\ket{1}$) in addition to the unknown coefficients ($\cos{(N_q\theta+{\pi}/{4})}$ and $\sin{(N_q\theta+{\pi}/{4})}$ with $\theta$ the parameter of interest); this differs from the usual setting of quantum metrology where only the coefficients are unknown~\cite{Giovannetti2005}.

The contribution of this paper is as follows. 
First, we prove that the quantum Fisher information matrix is diagonal in our problem setting, meaning that the QCRB of the amplitude parameter is not perturbed by the nuisance noise parameter. 
That is, fortunately, the above-mentioned first difficulty does not appear in our problem. 
Next, to deal with the second difficulty, we present a method for approaching the QCRB of the amplitude, using the so-called 2-step adaptive estimation strategy~\cite{Hayashi1998, Barndorff-Nielsen2000} combined with the variational quantum algorithm (VQA); more precisely, we first obtain a rough estimate of the amplitude using a measurement constructed via VQA to approximate the unknown optimal measurement, followed by performing a more precise estimation based on the rough estimate obtained in the first-step. 
Note that a similar technique was proposed in Ref.~\cite{Meyer2021}, which adjusts the measurement basis using VQA to approximate the optimal estimator.

This paper is organized as follows.
In Section~\ref{SEC_Preliminaries}, we introduce the maximum likelihood method for amplitude estimation, and present some basic notions of quantum estimation theory, including QCRB. 
Section~\ref{SEC_QCRB_optimal_measurement} presents our first main result; we prove that the quantum Fisher information matrix is diagonal in our case. 
In addition, we derive the optimal measurements for the noisy amplitude estimation problem. 
Section~\ref{SEC_Method} provides our second main result, showing the VQA-based method for attaining QCRB of the amplitude parameter. 
Section~\ref{SEC_Result} is devoted to numerical demonstrations of the proposed method.
We conclude this paper with Section~\ref{SEC_Conclusion}.

\section{Preliminaries}
\label{SEC_Preliminaries}


\subsection{Maximum likelihood amplitude estimation}
\label{SEC_MLAE}

The methodology presented in this paper is based on maximum likelihood amplitude estimation (MLAE)~\cite{Y.Suzuki2020}.
MLAE combines amplitude amplification with the maximum likelihood estimation (MLE). 
Below we give a summary of MLAE.

Suppose a single real-valued parameter $\theta$ is embedded in the amplitude of the following $(n+1)$-qubits state $\ket{\psi}_{n+1}$ by a black-box unitary operator $A$; 
\begin{equation}
    \label{EQ_psi}
    A\ket{0}_{n+1} = \ket{\psi(\theta)}_{n+1} 
      = \cos{\theta}\ket{\psi_0}_n\ket{0} + \sin{\theta}\ket{\psi_1}_n\ket{1},
\end{equation}
where $0 \le \theta \le \pi/2$. 
$\ket{\psi_0}_n$ and $\ket{\psi_1}_n$ are some $n$-qubits states. 
Note that $A$ is implementable but unknown due to the black-box assumption; thus, $\ket{\psi_0}_n$ and $\ket{\psi_1}_n$ are also unknown.
$\ket{0}$ and $\ket{1}$ are orthogonal single-qubit ancilla states. 
The objective of MLAE is to estimate the unknown parameter $\theta$.

In MLAE, we initially apply the amplitude amplification (Grover) operator $G$ on $\ket{\psi (\theta) }_{n+1}$.
Here, $G$ is defined as $G=AS_0A^{\dagger}S_{f}$, where $S_0=-I_{n+1}+2\ket{0}_{n+1}\bra{0}_{n+1}$ and $S_f=-I_{n+1}+2I_n \otimes \ket{0} \bra{0}$. 
$I_n$ is the identity operator for $n$-qubits. 
By applying the operator $G$ on state \eqref{EQ_psi} iteratively $m$ times, we have 
\begin{align}
    \label{EQ_psi_amplitude_amplification}
    G^{m}\ket{\psi(\theta)}_{n+1} &= \cos{((2m+1)\theta)}\ket{\psi_0}_n\ket{0} \nonumber \\
    &\hspace{11pt} + \sin{((2m+1)\theta)}\ket{\psi_1}_n\ket{1} \nonumber \\
    &=:\ket{\psi(N_q\theta)} ,
\end{align}
where $N_q = 2m + 1$. 
Then, the measurement is performed on the ancilla qubit of quantum state $G^{m}\ket{\psi(\theta)}_{n+1}$ in the computational basis. 
The probabilities of obtaining the measurement results 0 and 1 are respectively given by 
\begin{align}
    \label{EQ_probability}
    &{\rm Pr}(0 ; \theta, m) = \cos^2{(2m+1)\theta}, \nonumber \\
    &{\rm Pr}(1 ; \theta, m) = \sin^2{(2m+1)\theta}.
\end{align}
We now prepare a set of sequences $\{m_k\}_{k=0}^M$ and perform the measurement for each $m = m_k$.
When $N_{\rm shot}$ measurements are made and the result 0 is obtained $h_k$ times for each $m_k$, the total likelihood function $f^{\rm MLAE}_L$ is expressed as follows:
\begin{align}
    \label{EQ_loglikelihood_MLAE}
    f^{\rm MLAE}_L(\theta; {\bf h}) = \prod_{k=0}^M \left[ {\rm Pr}(0 ; \theta, m_k) \right]^{h_k} \left[ {\rm Pr}(1 ; \theta, m_k) \right]^{N_{\rm shot} - h_k} ,
\end{align}
where ${\bf h}=(h_0,h_1,...,h_M)$. 
Given ${\bf h}$ as the measurement results, the maximum likelihood estimator for $\theta$ is calculated as $\hat{\theta}_{\rm ML}={\rm argmax}_\theta f_L^{\mathrm{MLAE}}(\theta; {\bf h})$.

In general, any unbiased estimator $\hat{\theta}$ satisfies the Cram\'{e}r-Rao inequality~\cite{Rao1973}, which, in our case, is 
\begin{align}
    \label{EQ_CCRB_def}
    \mathbb{E}[(\hat{\theta} - \theta)^2] \ge \cfrac{1}{F_c} = \cfrac{1}{\sum_{k=0}^M F_c(m_k)},
\end{align}
where $F_c(m_k)$ is the Fisher information obtained by measuring the quantum state $\ket{\psi(N_q\theta)} = G^{m_k}\ket{\psi(\theta)}_{n+1}$. 
(Here, $N_q=2m_k+1$. For simplicity, we will maintain the same notation as $N_q=2m+1$.)
Specifically, from~\cite{Uno2021}, we have $F_c(m_k) = N_{\rm shot} [4(2m_k+1)^2]$ . 
Because $(\sum_{k=0}^{M}2m_k+1)^2 \ge \sum_{k=0}^{M}(2m_k+1)^2$, the following inequality holds
\begin{align}
    \label{EQ_CCRB}
    \mathbb{E}[(\hat{\theta} - \theta)^2] \ge \cfrac{1}{4 N_{\rm shot} \sum_{k=0}^{M}(2m_k+1)^2} \ge \cfrac{1}{4 N_{\rm shot} N_A^2},
\end{align}
where $N_A=\sum_{k=0}^{M}(2m_k+1)$. 
When the number of Grover operators $m_k$ is increased according to the exponentially incremental sequence (EIS), i.e., $m_k=0, 2^0, 2^1,\cdots$, the equality in the right inequality of Eq.~(\ref{EQ_CCRB}) is nearly satisfied~\cite{Y.Suzuki2020}.
In addition, the maximum likelihood estimator asymptotically saturates Cram\'{e}r-Rao bound in the left inequality of Eq.~(\ref{EQ_CCRB}).
Thus, the estimation error $\epsilon=\sqrt{\mathbb{E}[(\hat{\theta} - \theta)^2]}$ decreases as $\mathcal{O}(1/\sqrt{N_{\rm shot}}N_A)$. 
Now recall that $2m_k$ is the number of $A$ contained in the entire unitary $G^{m_k}$; thus, the total number of queries of $A$ in the MLAE algorithm is $N_{\rm shot} N_A$. 
Hence, for a fixed $N_{\rm shot}$, the lower bound in Eq.~\eqref{EQ_CCRB} is quadratically lower than in the case without Grover operation (i.e., the case $m_k=0~\forall k$), which is interpreted as the quantum speedup in the context of MLAE.

\subsection{Noisy MLAE}
\label{SEC_Noisy_MLAE}

In the process of non-ideal quantum computation, the quantum state can be disrupted by possibly unknown external noises. 
Because the MLE method is essentially a model-based estimation method, we need a model that incorporates the effects of noise into the quantum state and the corresponding measurement probability.
Here we follow the approach of Ref.~\cite{Tanaka2021}, which assumes depolarizing noise~\cite{Nielsen2002} acts on the quantum state each time the Grover operator $G$ is applied , with an unknown probability $1-p$. 
Here, $p$ can be interpreted as the noise intensity. 
Under this assumption, after applying the Grover operator $m_k$ times, the quantum state is given by
\begin{equation}
    \label{EQ_rho}
    \rho_{m_k}(\theta,p) = p^{m_k} \ket{\psi(N_q\theta)}\bra{\psi(N_q\theta)} + (1 - p^{m_k})\cfrac{I_{n+1}}{d},
\end{equation}
where $d=2^{n+1}$. 
Thus, our task is to estimate both $\theta$ and $p$. 
For this multi-parameters case, we can apply the same MLE method explained in Section~\ref{SEC_MLAE} using the 2-dimensional likelihood function $f^{\rm MLAE}_L(\theta, p; {\bf h})$ constructed from the measurement results of $\rho_{m_k}(\theta,p)$~\cite{Tanaka2021}.

\subsection{Quantum Fisher information matrix and Quantum Cram\'{e}r-Rao bound}
\label{SEC_QCRB}

In the estimation of the parameter embedded in a quantum state, there exists an ultimate precision limit called the QCRB, which does not depend on the measurement basis~\cite{Helstrom1968}. 
In general, when estimating multiple parameters ${\bm \theta}=(\theta_1,\theta_2,\cdots, \theta_{N_{\rm param}})$ embedded in a quantum state $\rho({\bm \theta})$, the following inequality holds for the variance of the unbiased estimate $\hat{\theta}_i$\cite{Helstrom1968, Holevo2011, Braunstein1994}:
\begin{equation}
    \label{EQ_def_QCRB_general}
    \mathbb{E}[(\hat{\theta}_i - \theta_i)^2] \ge \cfrac{1}{N_{\rm shot}}[F_q^{-1}]_{\theta_i,\theta_i},
\end{equation}
where $F_{q}$ is the quantum Fisher information matrix (QFIM) and $\left[F_q^{-1}\right]_{\theta_i,\theta_j}$ represents the $(i,j)$-th element of $F_{q}^{-1}$; here we assume that $F_{q}$ is invertible. 
$N_{\rm shot}$ represents the number of copies of $\rho({\bm \theta})$, which is the same as the number of measurements once the measurement scheme is specified.

There are several definitions of QFIM; here we employ the symmetric logarithmic derivative (SLD) QFIM~\cite{Helstrom1968}.
The matrix element of SLD QFIM $\left[F_q\right]_{\theta_i,\theta_j}$ is defined by
\begin{equation}
    \label{EQ_def_QFIM}
    \left[F_q\right]_{\theta_i,\theta_j} = \cfrac{1}{2}{\rm Tr} \left[ \rho({\bm \theta})(L_{\theta_i}L_{\theta_j} + L_{\theta_j}L_{\theta_i})\right],
\end{equation}
where $L_{\theta_i}$ is a Hermitian operator called the SLD operator of $\theta_i$ satisfying 
\begin{equation}
    \label{EQ_def_SLD}
    \cfrac{\partial \rho({\bm \theta})}{\partial \theta_i} = \cfrac{1}{2}\left[ \rho({\bm \theta}) L_{\theta_i} + L_{\theta_i} \rho({\bm \theta}) \right].
\end{equation}
It has been shown that the QCRB of $\theta_i$ in Eq.~(\ref{EQ_def_QCRB_general}) is attainable via the measurement that is obtained from the spectral decomposition of the following operator~\cite{J.Suzuki2020}:
\begin{align}
    \bar{L}_{\theta_i} = \sum_{j=1}^{N_{\rm param}} [F_q^{-1}]_{\theta_j, \theta_i} L_{\theta_j}.
\end{align}
However, in general, $\bar{L}_{\theta_i}$ depends on the unknown parameters ${\bm \theta}$, and thus we cannot implement the optimal measurement. 
To address this challenge, previous studies proposed an adaptive estimation strategy that performs multiple estimation iterations while refining the measurement based on the estimated values in each step~\cite{Nagaoka1989, Fujiwara2006}. 
It was also shown that QCRB of 1-parameter of interest is attainable by the 2-step estimation~\cite{Hayashi1998, Barndorff-Nielsen2000}.

In MLAE, the QCRB is attainable in the ideal noiseless case, but the general attainability in the presence of noise is unknown; for instance, it is not if the computational basis measurement is employed~\cite{Uno2021}. 
Refs.~\cite{Uno2021, Wada2022} studied the problem of estimating the single parameter $\theta$, under the assumption that the noise parameter $p$ is known; 
then the QCRB is reduced to 
\begin{equation}
    \label{EQ_QCRB_estimate_1param}
    \mathbb{E}[(\hat{\theta} - \theta)^2] \ge \cfrac{1}{N_{\rm shot}[F_q]_{\theta, \theta}}.
\end{equation}
In this paper, we consider estimating both $\theta$ and $p$. 
Due to the general inequality $[F_q^{-1}]_{\theta_i,\theta_i} \ge 1 / [F_q]_{\theta_i,\theta_i}$, when estimating both $p$ and $\theta$, the lower bound of estimation error of $\theta$ becomes bigger than the case when estimating only $\theta$ with known $p$; 
the amount of tightness of this inequality is not obvious, and this point will be addressed in Section~\ref{SEC_derivation_QFIM_QCRB}.

\section{Optimal error bound and measurement basis in noisy MLAE}
\label{SEC_QCRB_optimal_measurement}

In this section, we derive the QCRB for $\theta$ in the noisy MLAE problem with unknown parameter $p$ and compare the result to the case where $p$ is assumed to be known. 
We then derive the optimal measurement basis for attaining this bound.

\subsection{Derivation of QFIM and QCRB in noisy MLAE}
\label{SEC_derivation_QFIM_QCRB}

For the density matrix $\rho_{m_k}(\theta,p)$ given in Eq.~(\ref{EQ_rho}), the SLD operators $L_\theta$ and $L_p$, which satisfy Eq.~(\ref{EQ_def_SLD}), are calculated as follows:
\begin{align}
    &L_\theta = \cfrac{2dN_qp^{m_k}}{2+(d-2)p^{m_k}}
    \begin{bmatrix} 
      -\sin{2N_q\theta} & \cos{2N_q\theta} & 0 & \dots  & 0 \\
      \cos{2N_q\theta} & \sin{2N_q\theta} & 0 & \dots  & 0 \\
      0      & 0      & 0      & \dots & 0 \\
      \vdots & \vdots & \vdots & \ddots & \vdots \\
      0 & 0 & 0 & \dots  & 0 
    \end{bmatrix}, \label{EQ_SLD_theta} \\
    &L_p = \cfrac{d m_k p^{m_k-1}}{\{1+(d-1)p^{m_k}\}(1-p^{m_k})} \nonumber \\ 
    &\hspace{30pt} \times
    \begin{bmatrix} 
      \cos^2{N_q\theta} & \cos{N_q\theta}\sin{N_q\theta} & 0 & \dots  & 0 \\
      \cos{N_q\theta}\sin{N_q\theta} & \sin^2{N_q\theta} & 0 & \dots  & 0 \\
      0      & 0      & 0      & \dots & 0 \\
      \vdots & \vdots & \vdots & \ddots & \vdots \\
      0 & 0 & 0 & \dots  & 0 
    \end{bmatrix} \nonumber \\
    &\hspace{20pt} - \cfrac{m_kp^{m_k-1}}{1-p^{m_k}} I_{n+1}, \label{EQ_SLD_p}
\end{align}
where $p\neq 1$ is assumed (i.e., there must be noise). 
The basis in the top-left 2×2 block of $L_\theta$ and $L_p$ is $(\ket{\psi_0}_n\ket{0}, \ket{\psi_1}_n\ket{1})$, while the basis for all the remaining elements is $\ket{\Psi_i}$ with $i=2, \cdots, 2^{n+1}-1$. 
Here, $\ket{\Psi_i} \perp \ket{\psi_0}_{n}\ket{0}, \ket{\psi_1}_{n}\ket{1}$ for all $i$.
The detailed derivation is presented in Appendix~\ref{SEC_derivation}. 
Then, from Eq.~\eqref{EQ_def_QFIM} the SLD QFIM $F_{q}$ is calculated as
\begin{align}
    \label{EQ_QFIM}
    F_q 
    &=\hspace{-5pt}\
    \begin{bmatrix}
     [F_q]_{\theta,\theta} & [F_q]_{\theta,p} \\
     [F_q]_{p,\theta} & [F_q]_{p,p} \\
    \end{bmatrix} \nonumber \\
    &=\hspace{-5pt}\
    \begin{bmatrix}
    \cfrac{4dN_q^2p^{2m_k}}{2+(d-2)p^{m_k}} & 0 \\
    0 & \cfrac{m_k^2p^{2(m_k-1)}(d-1)}{(1-p^{m_k})\{1+(d-1)p^{m_k}\}} \\
    \end{bmatrix}.
\end{align}
Importantly, $F_q$ is already a diagonal matrix, so $[F_q^{-1}]_{\theta, \theta}=1/[F_q]_{\theta, \theta}$ holds; this means that even when $p$ is unknown, we have the same QCRB as that when $p$ is known.
Consequently, the estimation error $\mathbb{E}[(\hat{\theta} - \theta)^2]$ is lower bounded by
\begin{align}
    \label{EQ_QCRB}
    \mathbb{E}[(\hat{\theta} - \theta)^2] &\ge \cfrac{1}{N_{\rm shot}}[F_q^{-1}]_{\theta,\theta} 
    = \cfrac{1}{N_{\rm shot}[F_q]_{\theta,\theta}} \nonumber \\
    &= \cfrac{1}{N_{\rm shot}} \cdot \cfrac{2+(d-2)p^{m_k}}{4dN_q^2p^{2m_k}}.
\end{align}


\subsection{Optimal measurement basis}
\label{SEC_optimal_measurement_basis}

Here, we derive the optimal measurement basis needed to attain the lower bound in Eq.~(\ref{EQ_QCRB}). 
As explained in Section~\ref{SEC_QCRB}, the optimal measurement basis is obtained from the spectral decomposition of the operator $\bar{L}_{\theta}$~\cite{J.Suzuki2020}. 
In our case, since $F_q$ in Eq.~(\ref{EQ_QFIM}) is a diagonal matrix, $\bar{L}_{\theta} = [F_q^{-1}]_{\theta, \theta} L_\theta$.
Thus, the eigenvectors of $\bar{L}_{\theta}$ are equivalent to those of $L_\theta$, given by
\begin{align}
    \label{EQ_optimal_measurement_basis}
    \ket{\lambda_0(\theta)}_{n+1} = &\cos{\left(N_q\theta+\cfrac{\pi}{4}\right)}\ket{\psi_0}_n\ket{0} \nonumber \\
    &+\sin{\left(N_q\theta+\cfrac{\pi}{4}\right)}\ket{\psi_1}_n\ket{1} ,\nonumber \\
    \ket{\lambda_1(\theta)}_{n+1} = &-\sin{\left(N_q\theta+\cfrac{\pi}{4}\right)}\ket{\psi_0}_n\ket{0} \nonumber \\
    &+\cos{\left(N_q\theta+\cfrac{\pi}{4}\right)}\ket{\psi_1}_n\ket{1} ,
\end{align}
and $\ket{\lambda_l}_{n+1}$ for $2 \leq l \leq 2^{n+1}-1$; they are all orthogonal, i.e., $\ket{\lambda_i}_{n+1} \perp \ket{\lambda_{j}}_{n+1}$ for all $i\neq j$.

\section{Adaptive MLAE method}
\label{SEC_Method}

\begin{figure*}[t]
    \includegraphics[width=2\columnwidth]{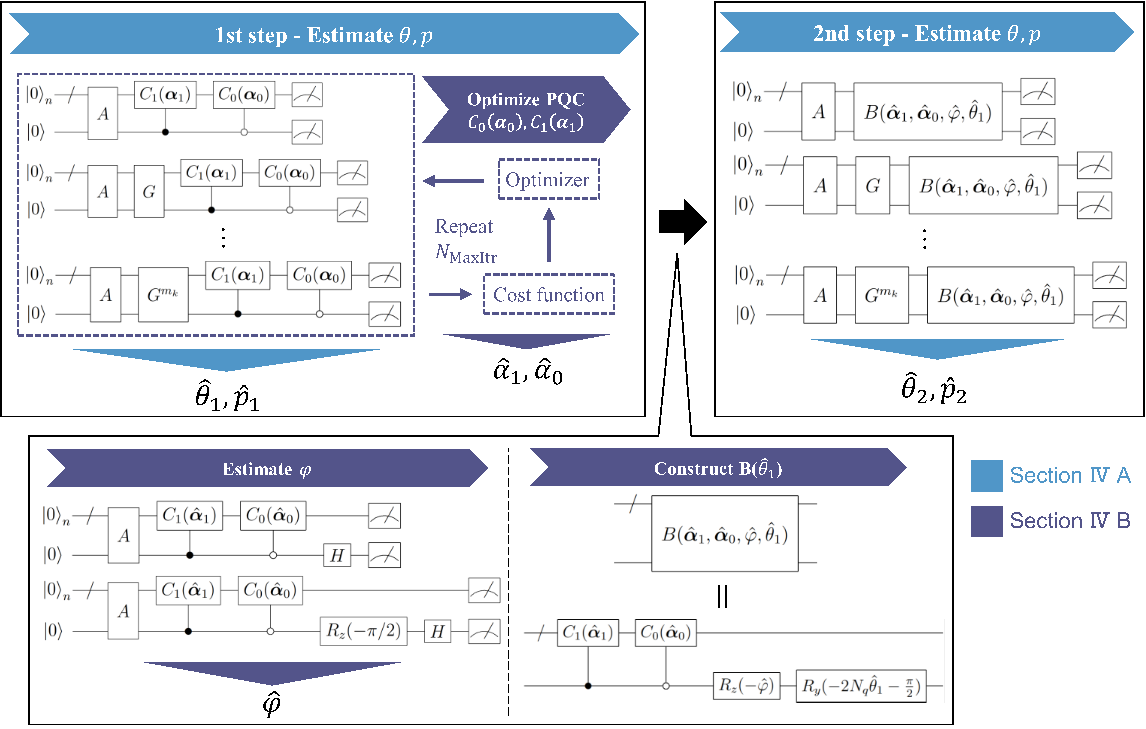}
    \caption{\label{Fig_algorithm_overview} 
    The overview of the proposed method.
    The entire procedure is divided into two parts: the adaptive measurement strategy explained in Section~\ref{SEC_adaptive}(blue) and the composition of the operator $B$ which adjusts the measurement basis explained in Section~\ref{SEC_optimize_vqc}(violet).
    Initially, we do both the first-step estimation of $\theta$, $p$ and the optimization of the variational quantum circuits $C_0({\bm \alpha}_0)$ and $C_1({\bm \alpha}_1)$ contained in $B$.
    In this step, the number of measurements (i.e. shots) for each circuit are $\sqrt{N_{\rm shot}}$.
    Next, we estimate the relative phase caused by the operation of $C_0({\bm \alpha}_0)$ and $C_1({\bm \alpha}_1)$ with $N_{\varphi - \rm shot}$ shots, and add the phase shift operation which cancels the relative phase.
    With the variational quantum circuits $C_0({\bm \alpha}_0), C_1({\bm \alpha}_1)$ and $R_z(-\hat{\varphi})$, we compose the operator B.
    Finally, we execute the second-step estimation of $\theta, p$ with $N_{\rm shot} - \sqrt{N_{\rm shot}}$ shots.
    In this second step, we optimize the measurement basis with operator $B$.
    }
\end{figure*}

Although we provided the explicit expression of the optimal measurement basis in the previous section, implementing these bases is generally impossible due to two factors:
(a) $\ket{\lambda_0(\theta)}_{n+1}$ and $\ket{\lambda_1(\theta)}_{n+1}$ depend on the unknown parameter $\theta$, and
(b) $\ket{\lambda_0(\theta)}_{n+1}$ and $\ket{\lambda_1(\theta)}_{n+1}$ depend on the unknown states $\ket{\psi_0}_n$ and $\ket{\psi_1}_n$. 
In the following, we outline the strategies for addressing (a) in Section~\ref{SEC_adaptive} and (b) in Section~\ref{SEC_optimize_vqc}.
An overview of the proposed strategy is illustrated in Fig. \ref{Fig_algorithm_overview}, and the pseudo code of the entire algorithm is described in Algorithm~\ref{ALG_whole}.

\subsection{Adaptive measurement strategy for MLAE}
\label{SEC_adaptive}

We apply the 2-step adaptive strategy~\cite{Hayashi1998, Barndorff-Nielsen2000} mentioned in Section~\ref{SEC_QCRB} to MLAE.
In the proposed method, first, we perform $\sqrt{N_{\rm shot}}$ measurements for each $m_k$ with the computational basis on each circuit, varying the number of operations of $G$.
Then, we obtain the initial rough estimates of $\hat{\theta}_1$ and $\hat{p}_1$.
Second, we optimize the measurement basis using the rough estimate $\hat{\theta}_1$
and perform $N_{\rm shot} - \sqrt{N_{\rm shot}}$ measurements for each $m_k$ to construct the improved estimates $\hat{\theta}_2$ and $\hat{p}_2$. 
We will numerically demonstrate that the estimate $\hat{\theta}_2$ asymptotically attains the lower bound of inequality (\ref{EQ_QCRB}). 

\begin{algorithm}[H]
    \caption{2-step adaptive MLAE}
    \label{ALG_adaptive}
    \begin{algorithmic}[1]
    \Require The operator $A$, the total number of measurements  \par
            $N_{\rm shot}$, the iteration limit of the Grover operator $M-1$
    \Ensure The estimate $\hat{\theta}_2$
    \State Construct $G$ with $A$
    \State $GroverList = {\{0,2^0,2^1,2^2,\cdots,2^{M-1}\}} $
    \For{$m_k$ in $GroverList$}
    \State $i =$ index of $m_k$ in $GroverList$
    \State Operate $A$ and $G^{m_k}$ on the initial state $\ket{0}_{n+1}$
    \State Perform $\sqrt{N_{\rm shot}}$ measurements with the computa-\par
           \hspace{-5pt}tional basis on the ancilla qubit, and retrieve the \par
           \hspace{-5pt}results $h_{1, i}$
    \EndFor
    \State Obtain $\hat{\theta}_1$ and $\hat{p}_1$ by MLE with ${\bf h}_1=(h_{1, 0},h_{1, 1}, \cdots ,h_{1, M})$
    \For{$m_k$ in $GroverList$}
    \State $i =$ index of $m_k$ in $GroverList$
    \State Operate $A$ and $G^{m_k}$ on the initial state $\ket{0}_{n+1}$
    \State Perform $N_{\rm shot}-\sqrt{N_{\rm shot}}$ measurements with the \par
           \hspace{-5pt}optimized basis \eqref{optimal POVM}, and retrieve the results ${\bf h}_{2, i}$
    \EndFor
    \State Obtain $\hat{\theta}_2$ and $\hat{p}_2$ by MLE with ${\bf h}_2=({\bf h}_{2, 0},{\bf h}_{2, 1}, \cdots ,{\bf h}_{2, M})$
    \State \Return $\hat{\theta}_2$
    \end{algorithmic}
\end{algorithm}

The pseudo-code of the above-described procedure is given in Algorithm \ref{ALG_adaptive}. 
In the first part (from Line 1 to 7 in Algorithm \ref{ALG_adaptive}), we perform the computational-basis measurement on the ancilla qubit after applying $G^{m_k}$ to the initial state. 
$h_{1,k}$ denotes the number of 0 obtained in this measurement process; 
the measurement data is used to construct the maximum likelihood estimator $(\hat{\theta}_1, \hat{p}_1)$. 
In the second part (from Line 8 to 13 in Algorithm \ref{ALG_adaptive}), we use $\hat{\theta}_1$ to construct a measurement that approximates the optimal one. 
Naively, this can be done via the projection measurement onto $\ket{\lambda_0(\hat{\theta}_1)}_{n+1}$, $\ket{\lambda_1(\hat{\theta}_1)}_{n+1}$, or the set of other states $\{\ket{\lambda_l}_{n+1}\}_{l=2}^{2^{n+1}-1}$; that is, the measurement with the projection operators
\begin{align}
\label{optimal POVM}
   \ket{\lambda_0(\hat{\theta}_1)}_{n+1}&\bra{\lambda_0(\hat{\theta}_1)},~~
   \ket{\lambda_1(\hat{\theta}_1)}_{n+1}\bra{\lambda_1(\hat{\theta}_1)}, \nonumber \\
   I-\ket{\lambda_0(\hat{\theta}_1)}_{n+1}&\bra{\lambda_0(\hat{\theta}_1)}
    - \ket{\lambda_1(\hat{\theta}_1)}_{n+1}\bra{\lambda_1(\hat{\theta}_1)}.
\end{align}
However, these measurement bases contain unknown states $\ket{\psi_0}_n$ and 
$\ket{\psi_1}_n$, and thus implementing this measurement is not possible even if $\theta$ is specified. 
Yet, we here assume that a perfect measurement~\eqref{optimal POVM} can be ideally given; a concrete procedure for constructing a measurement that approximates the optimal one~\eqref{optimal POVM} will be addressed in the next subsection. 
Under this ideal assumption, the probabilities of the measurement result of $\rho_{m_k}(\theta,p)$ in Eq.~(\ref{EQ_rho}) are given by
\begin{align}
    \label{EQ_probability_optbasis}
    {\rm Pr}(\lambda_0 ; \theta, p, m_k) = \; & p^{m_k} \cfrac{1 + \sin{2N_q \left(\theta - \hat{\theta}_1 \right)}}{2} +\frac{1-p^{m_k}}{d}, \nonumber \\
    {\rm Pr}(\lambda_1 ; \theta, p, m_k) = \; & p^{m_k} \cfrac{1 - \sin{2N_q \left(\theta - \hat{\theta}_1 \right)}}{2} +\frac{1-p^{m_k}}{d}, \nonumber \\
    {\rm Pr}(\lambda_l ; \theta, p, m_k) = \; & \cfrac{(1-p^{m_k})}{d}(d-2). 
\end{align}
That is, ${\rm Pr}(\lambda_0), {\rm Pr}(\lambda_1), {\rm Pr}(\lambda_l)$ represent the probability of the result corresponding to the measurement bases $\ket{\lambda_0(\theta)}_{n+1},\ket{\lambda_1(\theta)}_{n+1}, \{\ket{\lambda_l}_{n+1}\}$, respectively. 
We repeat this measurement process; in Line 12 of Algorithm~\ref{ALG_adaptive}, each element of ${\bf h}_{2, k} = (h_{2, k, \lambda_0}, h_{2, k, \lambda_1}, h_{2, k, \lambda_{l}})$ denotes the count of the measurement results corresponding to one of the above three measurement bases. 
Then, we construct the likelihood function using the same method introduced in Section~\ref{SEC_MLAE}, which is given by
\begin{align}
    \label{EQ_loglikelihood}
    f_L^{\mathrm{MLAE}}(\theta, p; {\bf h}_2) = \prod_{k=0}^M 
                    &\left[ {\rm Pr}(\lambda_0 ; \theta, p, m_k) \right]^{h_{2,k,\lambda_0}} \nonumber \\
               \cdot&\left[ {\rm Pr}(\lambda_1 ; \theta, p, m_k) \right]^{h_{2,k,\lambda_1}} \nonumber \\
               \cdot&\left[ {\rm Pr}(\lambda_l ; \theta, p, m_k) \right]^{h_{2,k,\lambda_l}}.
\end{align}
Finally, we obtain the estimates $(\hat{\theta}_2, \hat{p}_2)$ as the values of $(\theta, p)$ that maximize $f_L^{\mathrm{MLAE}}(\theta, p; {\bf h}_2)$.


\subsection{Variational search for the measurement basis}
\label{SEC_optimize_vqc}

\begin{figure}[H]
    \centering
    \subfigure[]
    {\includegraphics{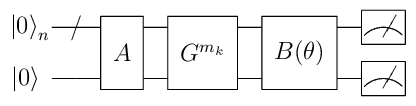}}
    \subfigure[]{\includegraphics[scale=0.8]{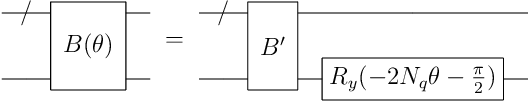}}
    \subfigure[]
    {\includegraphics[scale=0.8]{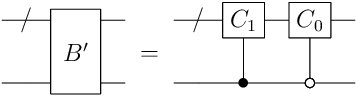}}
\caption{\label{Fig_circuit_AGB_overview} (a)The quantum circuit for MLAE with adjusting measurement basis by the operator $B(\theta)$, and the decomposition of (b) the operator $B(\theta)$ and (c) the operator $B'$.}
\end{figure}

As mentioned above, the optimal measurement \eqref{optimal POVM} cannot be implemented as they contain the unknown states $\ket{\psi_0}_n$ and 
$\ket{\psi_1}_n$. 
Here, we propose a method of using a variational quantum circuit (VQC) that approximates the optimal measurement. 
Specifically, we propose adjusting the measurement basis using the operator $B(\theta)$, as illustrated in Fig.~\ref{Fig_circuit_AGB_overview}(a). 
In typical quantum computing scenarios, we can only perform a measurement in a computational basis. 
Hence, $B(\theta)$ is used to transform the optimal basis (\ref{EQ_optimal_measurement_basis}) to the computational basis. 
As a simple choice, we take 
\begin{align}
\label{EQ_operator_B}
    B(\theta)\ket{\lambda_0(\theta)}_{n+1} = \ket{0}_n\ket{0}, \nonumber \\
    B(\theta)\ket{\lambda_1(\theta)}_{n+1} = \ket{0}_n\ket{1}.
\end{align}
In this case, $\{\ket{\lambda_l}_{n+1}\}$ represents all computational bases orthogonal to $\ket{0}_n\ket{0}$ and $\ket{0}_n\ket{1}$.

Moreover, we decompose the operation (\ref{EQ_operator_B}) to $B'$ and $R_y(-2N_q\theta-\frac{\pi}{2})$, as illustrated in Fig. \ref{Fig_circuit_AGB_overview}(b). 
First, we apply the operator $B'$, which eliminates the unknown states as follows:
\begin{align}
    \label{EQ_operator_B'}
    &B' \ket{\lambda_0(\theta)}_{n+1} = \cos{\left(N_q\theta+\cfrac{\pi}{4}\right)}\ket{0}_n\ket{0} \nonumber \\
    &\hspace{50pt} +\sin{\left(N_q\theta+\cfrac{\pi}{4}\right)}\ket{0}_n\ket{1} ,\nonumber \\
    &B' \ket{\lambda_1(\theta)}_{n+1} = -\sin{\left(N_q\theta+\cfrac{\pi}{4}\right)}\ket{0}_n\ket{0} \nonumber \\
    &\hspace{50pt} +\cos{\left(N_q\theta+\cfrac{\pi}{4}\right)}\ket{0}_n\ket{1}.
\end{align}
Next, we apply $R_y(-2N_q\theta-\frac{\pi}{2})$, the rotation operation around the $y$-axis, to the ancilla qubit.
With these operations, we can decompose $B(\theta)$ into a $\theta$-dependent component $R_y(-2N_q\theta-\frac{\pi}{2})$ and a $\theta$-independent component $B'$.


Here, we present a procedure for composing $B'$. 
Note again that $B'$ itself contains the unknown quantum states $\ket{\psi_0}_n$ and $\ket{\psi_1}_n$, meaning that a perfect construction of $B'$ satisfying all the requirements is not feasible. 
To address this issue, we divide $B'$ into two operators, $C_0$ and $C_1$ as shown in Fig. \ref{Fig_circuit_AGB_overview}(c) such that
\begin{align}
    \label{EQ_operator_C0C1}
    C_0\ket{\psi_0}_n = \ket{0}_n, ~~
    C_1\ket{\psi_1}_n = \ket{0}_n.
\end{align}
Next, we approximate $C_0$ and $C_1$ by VQCs, denoted as $C_0({\bm \alpha}_0)$ and $C_1({\bm \alpha}_1)$ respectively, using variational parameters ${\bm \alpha}_0$ and ${\bm \alpha}_1$.
Since $\ket{\psi_0}_n$ and $\ket{\psi_1}_n$ are generally different states, we need to operate $C_0({\bm \alpha}_0)$ and $C_1({\bm \alpha}_1)$ controlled by the ancilla qubit.
Each VQC comprises multiple layers of parametrized $R_y$ and $R_z$ gates and the controlled-$X$ gates between two adjacent qubits. 
We assume that implementation cost of $B'$ is sufficiently lower than that of $A$ and can be ignored. 
This allows us to assess the query complexity of the proposed algorithm by focusing solely on $A$, and the total hardware noise on $B'$ can be ignored under this assumption.
Since $A$ generally consists of numerous controlled operations, this assumption might be reasonable.
Note that even if implementation cost of $B'$ is too large to be ignored, as long as it is no greater than that of $A$, the framework of this paper remains applicable with a minor extension.

\begin{figure}[H]
    \centering
    \includegraphics[width=\columnwidth]{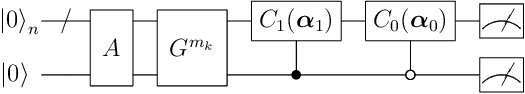}
    \caption{\label{Fig_circuit_AGC} The quantum circuit for optimizing ${\bm \alpha}_0$ and ${\bm \alpha}_1$.}
\end{figure}

To optimize the variational parameters ${\bm \alpha}_0$ and ${\bm \alpha}_1$, we employ the measurement results obtained from the circuit in Fig.~\ref{Fig_circuit_AGC}, utilizing the EIS sequence $\{m_k\}_{k=0}^{M} = \{0, 2^0, 2^1,..., 2^{M-1}\}$.
Our objective is to minimize the following cost function $f_{\rm cost}$, to ensure that $C_0({\bm \alpha}_0)$ and $C_1({\bm \alpha}_1)$ perform the desired operation, as specified in Eq.~(\ref{EQ_operator_C0C1}).
The cost function is defined as follows:
\begin{align}
    \label{EQ_cost_function}
    f_{\rm cost}({\bm \alpha}_0,{\bm \alpha}_1) &= \sum_{m_k} {\rm Tr} \left[ O_{\rm L} B'({\bm \alpha}_0, {\bm \alpha}_1) \ket{\psi((2m_k+1)\theta)} \right. \nonumber \\
    &\hspace{45pt} \left. \bra{\psi((2m_k+1)\theta)}B'^\dagger({\bm \alpha}_0, {\bm \alpha}_1) \right], \nonumber \\
    O_{\rm L} &= I_{n+1}  - \cfrac{1}{n}\sum_{j=1}^n \ket{0}\bra{0}_j \otimes I_{\Bar{j}},
\end{align}
where $I_{\Bar{j}}$ denotes the identity operator on all qubits except for the $j$-th qubit.
Same as the cost function proposed in~\cite{Khatri2019}, $f_{\rm cost}$ counts the case where the measurement results of each qubit, excluding the ancilla qubit, are not $\ket{0}$.
We can estimate the gradients of $f_{\rm cost}$ for each element $\alpha_k$ of ${\bm \alpha}_0$ and ${\bm \alpha}_1$ using the parameter-shift rule~\cite{Mitarai2018, Schuld2019} with the measurement result, and update the value of ${\bm \alpha}_0$ and ${\bm \alpha}_1$ by the gradient descent method. 

Note that, in general, optimization of VQC suffers from the so-called barren plateau issue, wherein the gradients of variational parameters diminish exponentially with the number of qubits~\cite{McClean2018}. 
Previous research has shown that using a shallow depth circuit and a local cost function, characterized solely by single qubit measurement results, is an efficient way to circumvent this obstacle~\cite{Cerezo2021}.
Our cost function (\ref{EQ_cost_function}) evaluates the measurement results of each individual qubit, meaning that Eq.~(\ref{EQ_cost_function}) is a local cost; 
additionally, we use a shallow VQC. 
Hence, it is presumed that the barren plateau could be avoided, and actually a supporting numerical simulation result will be demonstrated in Section~\ref{SEC_Result_vqc_var-grad}.

It is notable that the measurement results used for optimizing ${\bm \alpha}_0$ and ${\bm \alpha}_1$, can also be used for estimating $\theta$ and $p$.
In the optimization procedure, the ancilla qubit in Fig.~\ref{Fig_circuit_AGC} functions as the control qubit.
This means that the probability of the measurement results of the ancilla qubit is the same as Eq.~(\ref{EQ_probability}).
Therefore, during the optimization process of ${\bm \alpha}_0$ and ${\bm \alpha}_1$ via the measurement on the ancilla qubit, we can run the method explained in Section~\ref{SEC_MLAE} and Section~\ref{SEC_Noisy_MLAE} to estimate $\theta$ and $p$. 

With the above procedure, we can approximate the operator $B'$ by $B'(\hat{\bm \alpha}_0, \hat{\bm \alpha}_1)$, where $\hat{\bm \alpha}_0$ and $\hat{\bm \alpha}_1$ are the optimized values of ${\bm \alpha}_0$ and ${\bm \alpha}_1$.
However, even if $f_{\rm cost}(\hat{\bm \alpha}_0, \hat{\bm \alpha}_1)=0$, the operation of $B'(\hat{\bm \alpha}_0, \hat{\bm \alpha}_1)$ can cause the relative phase $\varphi$, compared to the ideal operation (\ref{EQ_operator_B'})
\begin{align}
    \label{EQ_B'_phase}
    B'(\hat{\bm \alpha}_0, \hat{\bm \alpha}_1)&\ket{\psi(N_q\theta)}_{n+1} \nonumber \\
     = &\cos{N_q\theta}\ket{0}_n\ket{0} + e^{i\varphi}\sin{N_q\theta}\ket{0}_n\ket{1}.
\end{align}
This is because our cost function (\ref{EQ_cost_function}) cannot distinguish the phase differences.
Therefore, it is necessary to estimate the relative phase $\varphi$ and apply the phase shift operation with  $R_z(-\hat{\varphi})$, which is the rotation operation around the $z$-axis, constructed using the estimate $\hat{\varphi}$ as depicted in Fig.~\ref{Fig_circuit_B_C_phi}. 
We show the detailed method for estimating $\varphi$ in Appendix~\ref{SEC_estimate_varphi}.
\begin{figure}[H]
    \centering
    \includegraphics[width=\columnwidth]{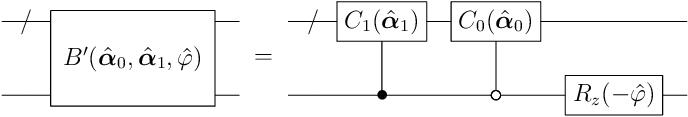}
    \caption{\label{Fig_circuit_B_C_phi} The construction of $B'(\hat{\bm \alpha}_0, \hat{\bm \alpha}_1, \hat{\varphi})$.}
\end{figure}

Algorithm~\ref{ALG_whole} represents the pseudo-code of the entire procedure, which modifies the Algorithm~\ref{ALG_adaptive} with VQC explained above.
Algorithm~\ref{ALG_VQA_MLAE} describes the procedure for executing MLAE and optimization of VQC simultaneously and is called as a subroutine in Algorithm~\ref{ALG_whole} .
Here, $N_{\rm MaxItr}$ denotes the iteration limit of the optimization of the variational parameters ${\bm \alpha}_0, {\bm \alpha}_1$.
\begin{algorithm}[H]
    \caption{2-step adaptive MLAE with VQC}
    \label{ALG_whole}
    \begin{algorithmic}[1]
    \Require The operator $A$, the total number of measurements  \par
            $N_{\rm shot}$, the iteration limit of the Grover operator \par
            $M-1$, the operator $C_0({\bm \alpha}_0)$ and $C_1({\bm \alpha}_1)$ constructed \par
            with VQC, the iteration limit of the optimization \par
            $N_{\rm MaxItr}$
    \Ensure The estimate $\hat{\theta}_2$
    \State Construct $G$ with $A$
    \State $GroverList = {\{0,2^0,2^1,2^2,\cdots,2^{M-1}\}} $
    \State Obtain $\hat{\theta}_1$, $\hat{p}_1$, $\hat{\bm \alpha}_0$ and $\hat{\bm \alpha}_1$ by Algorithm \ref{ALG_VQA_MLAE}, with $A$, $G$, $GroverList$, $\sqrt{N_{\rm shot}}$, $C_0({\bm \alpha}_0)$, $C_1({\bm \alpha}_1)$, and $N_{\rm MaxItr}$
    \State Estimate $\varphi$ with procedure in Appenix B
    \State Compose $B'(\hat{\bm \alpha}_0, \hat{\bm \alpha}_1, \hat{\varphi})$ represented in Fig.~\ref{Fig_circuit_B_C_phi}
    \State Compose $B(\hat{\theta}_1, \hat{\bm \alpha}_0, \hat{\bm \alpha}_1, \hat{\varphi})$ represented in Fig.~\ref{Fig_circuit_AGB_overview} (b) (replace $B'$ with $B'(\hat{\bm \alpha}_0, \hat{\bm \alpha}_1, \hat{\varphi})$)
    \State Obtain $\hat{\theta}_2$ and $\hat{p}_2$ by line 9 to 13 of Algorithm \ref{ALG_adaptive}, with $A$, $G$, $GroverList$, $N_{\rm shot} - \sqrt{N_{\rm shot}}$, and $B(\hat{\theta}_1)$, while optimizing measurement basis with $B(\hat{\theta}_1)$
    \State \Return $\hat{\theta}_2$
    \end{algorithmic}
\end{algorithm}

\begin{algorithm}[H]
    \caption{Simultaneous execution of MLAE and optimization of VQC}
    \label{ALG_VQA_MLAE}
    \begin{algorithmic}[1]
    \Require The operator $A$, the Grover operator $G$, the Grover iteration list $GroverList$, the total number of measurements $N'_{\rm shot}$ ,the operator $C_0({\bm \alpha}_0)$ and $C_1({\bm \alpha}_1)$ constructed with VQC, the iteration limit of the optimization $N_{\rm MaxItr}$
    \Ensure The estimates $\hat{\theta}$, $\hat{p}$, $\hat{\bm \alpha}_0$ and $\hat{\bm \alpha}_1$
    \State Initialize ${\bm \alpha}_0, {\bm \alpha}_1$
    \For{$m_k$ in $GroverList$}
    \State $i =$ index of $m_k$ in $GroverList$
    \For {$j=1$ to $N_{\rm MaxItr}$}
    \State Perform $N'_{\rm shot} / N_{\rm MaxItr}$ measurements on the quan-\par
        \hspace{10pt}tum circuit in Fig.~\ref{Fig_circuit_AGC}
    \State Calculate $f_{\rm cost}$ from the measurement results
    \State Calculate $\left\{ {\partial f_{\rm cost}}/{\partial\alpha_k} \right\}_{\forall \alpha_k}$
    \State Update ${\bm \alpha}_0, {\bm \alpha}_1$ with $\left\{ {\partial f_{\rm cost}}/{\partial\alpha_k} \right\}_{\forall \alpha_k}$
    \EndFor
    \State Combine the measurement results of ancilla qubit \par
        \hspace{-5pt}into $h_{i}$
    \EndFor
    \State Obtain $\hat{\theta}$ and $\hat{p}$ by MLE with ${\bf h}=(h_{0},h_{1}, \cdots ,h_{M})$  
    \State \Return $\hat{\theta}$, $\hat{p}$, $\hat{\bm \alpha}_0$ and $\hat{\bm \alpha}_1$
    \end{algorithmic}
\end{algorithm}

Note that, in this methodology, the variational optimization of ${\bm \alpha}_0$ and ${\bm \alpha}_1$ can be imperfect. 
In such a case, $C_0(\hat{\bm \alpha}_0)$ and $C_1(\hat{\bm \alpha}_1)$ perform 
\begin{align}
    \label{EQ_operator_C0C1_error}
    &C_0(\hat{\bm \alpha}_0) \ket{\psi_0}_n = p_{c_0}\ket{0}_n + \sum^{2^n-1}_{j=1}p_{c_0,j}'\ket{j}_n \nonumber \\
    &C_1(\hat{\bm \alpha}_1) \ket{\psi_1}_n = p_{c_1}\ket{0}_n + \sum^{2^n-1}_{j=1}p_{c_1,j}'\ket{j}_n \nonumber \\
    {\rm where \quad} &|p_{c_0}|^2+\sum^{2^n-1}_{j=1}|p_{c_0,j}'|^2=1 , \quad |p_{c_1}|^2+\sum^{2^n-1}_{j=1}|p_{c_1,j}'|^2=1.
\end{align}
Here, $p_{c_0}$ and $p_{c_1}$ represent the degree to which the operation of $C_0(\hat{\bm \alpha}_0)$ and $C_1(\hat{\bm \alpha}_1)$ correspond to the ideal ones. 
In the following, we summarize the effect of $p_{c_0}, p_{c_1}$ on the estimation performance, the detailed analysis of which is presented in Appendix~\ref{SEC_bias}.

Initially, we consider the effect of $p_c$, which combine the effect of $C_0(\hat{\bm \alpha}_0)$ and $C_1(\hat{\bm \alpha}_1)$.
Here, $p_c$ is defined as $|p_c|^2 \equiv |p_{c_0}|^2\cos^2{N_q\theta} + |p_{c_1}|^2\sin^2{N_q\theta}$.
As explained in Appendix~\ref{SEC_bias}, $p_c$ has a similar effect to the noise parameter $p$.
Nevertheless, the effect of $p_c$ remains constant regardless of the number of Grover operations $m_k$ and is negligible compared to the effect of depolarizing noise when $m_k$ is large.

In addition, depending on the value of $|p_{c_0}| - |p_{c_1}|$, the estimation bias $\theta_p$ arises. 
When $\theta_p$ is not negligible, we must simultaneously estimate $p_{c_0}$ and $p_{c_1}$ along with $\theta$ to assess $\theta_p$.
In Appendix \ref{SEC_estimate_bias}, we demonstrate this simultaneous estimation of $\theta, p, p_{c_0},$ and $p_{c_1}$, and show that the estimation error is conserved when $|p_{c_0}| - |p_{c_1}|$ is sufficiently small.

\section{Numerical simulation}
\label{SEC_Result}

We assess the effectiveness of the proposed method through numerical experiments conducted on a classical simulator.
In Section~\ref{SEC_Result_vqc_var-grad}, we evaluate the trainability of the variational circuit $C_0({\bm \alpha}_0)$ and $C_1({\bm \alpha}_1)$.
Section~\ref{SEC_Result_adaptive-estimation} demonstrates that the estimator can attain nearly the optimal QCRB.

\subsection{Trainability of the variational quantum circuit}
\label{SEC_Result_vqc_var-grad}

To examine whether the variational method proposed in Section~\ref{SEC_optimize_vqc} effectively mitigates the barren plateau issue, we evaluate the variance of the gradient of cost function, for several system sizes (the number of qubits $n$ and the layers of VQC $N_L$).
In the calculation, we focus on a single parameter of VQC, while other parameters are randomly chosen. 
We calculate the gradient $N_{\rm sample}=300$ times using the parameter shift rule while changing the values set with these random numbers; then we compute the variance of the gradient.

\begin{figure}[h]
    \centering
    \includegraphics[width=\columnwidth]{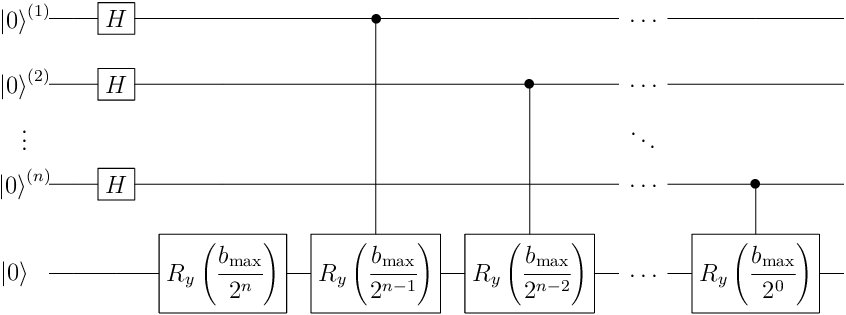}
    \caption{\label{Fig_circuit_A} The composition of the operator $A_{\rm sin}$ defined in Eq.~(\ref{EQ_operator_A}).}
\end{figure}

As an example, the operator $A$ is chosen as 
\begin{align}
    \label{EQ_operator_A}
    &A_{\rm sin}\ket{0}_{n+1} =  \sqrt{1 - S} \ket{\psi_0}_n \ket{0} + \sqrt{S} \ket{\psi_1}_n \ket{1} \nonumber \\
    &S = \sum^{2^n-1}_{x=0} \cfrac{1}{2^n} \sin^2{\left( \cfrac{(x+\frac{1}{2})b_{\rm max}}{2^n} \right)}.
\end{align}
This operator $A_{\rm sin}$ could be implemented efficiently in the circuit illustrated in Fig. \ref{Fig_circuit_A}~\cite{Y.Suzuki2020}.
In this evaluation, we set $b_{\rm max}=1/4$. 
Note that, in this setting, the amplitude of $\ket{\psi_0}_n$ and $\ket{\psi_1}_n$ are restricted to real values.
Thus the VQC used in this simulation does not include $R_z$ gates.
The other parameters are summarized in Table~\ref{TABLE_var-grad}.

\begin{table}[H]
    \caption{List of parameters for the numerical simulation of Section~\ref{SEC_Result_vqc_var-grad}}
    \label{TABLE_var-grad}
    \centering
    \begin{ruledtabular}
    \begin{tabular}{cll}
        number of measurements  & $N_{\rm shot}$  & $\infty~(\rm statevector)$ \\
        number of qubits  & $n$  & $\{4,6,8,10,12\}$ \\
        number of layers  & $N_L$   & $\{4,6,8,10,12,14\}$ \\
        number of samples\hspace{8pt}  & $N_{\rm sample}$  & 300 \\
    \end{tabular}
    \end{ruledtabular}
\end{table}

\begin{figure}[h]
    \centering
    \includegraphics[width=\columnwidth]{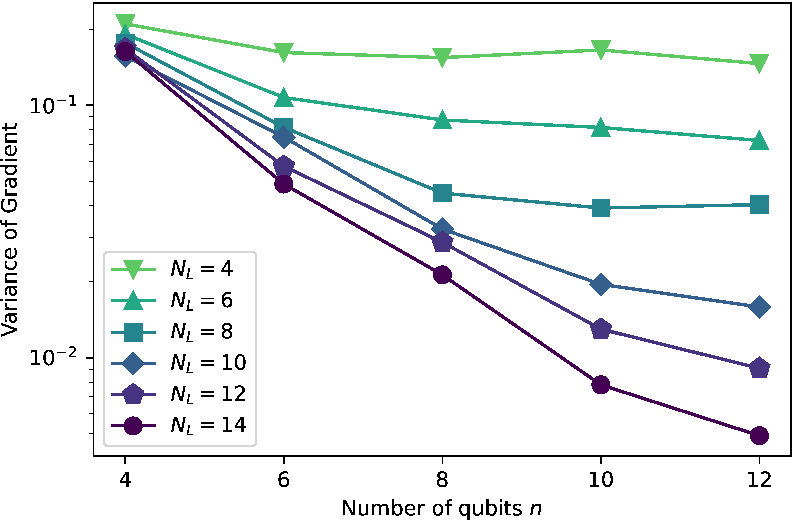}
    \caption{\label{Fig_result_var-grad} The relation of the system size(the number of qubits $n$ and the layer of VQC $N_L$) and the variance of the cost function gradient.}
\end{figure}

Figure \ref{Fig_result_var-grad} shows that, while the number of qubits and layers of the VQC increase, the variance of the cost function gradient does not decay exponentially fast with respect to $n$. 
That is, the variational optimization of ${\bm \alpha}_0$ and ${\bm \alpha}_1$ does not largely suffer from the barren plateau issue, thanks to the locality of the cost function and the shallowness of the VQC.

\subsection{Validation of Adaptive strategy}
\label{SEC_Result_adaptive-estimation}
\begin{figure*}[t]
    \centering
    \subfigure[]{\includegraphics[width=0.99\columnwidth]{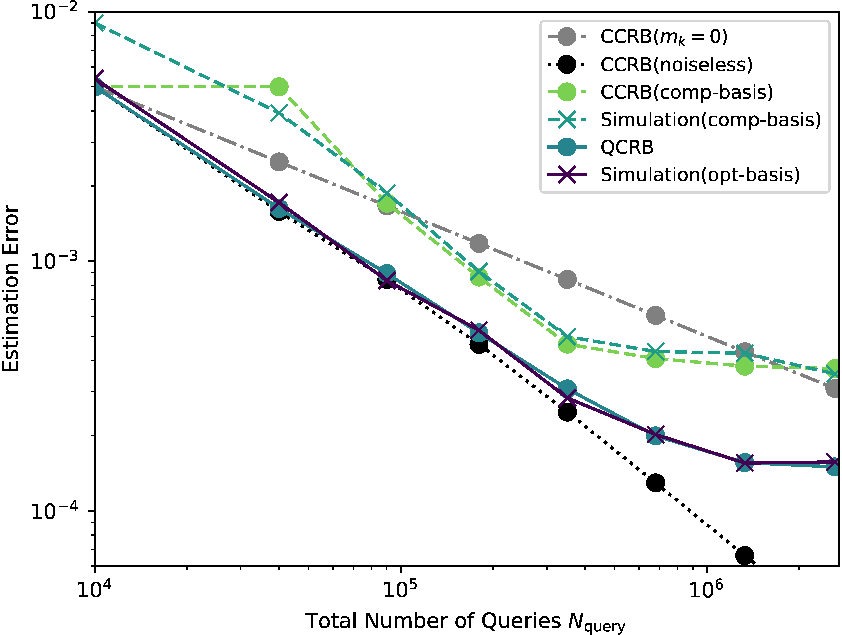}} \hspace{0.02\columnwidth}
    \subfigure[]{\includegraphics[width=0.99\columnwidth]{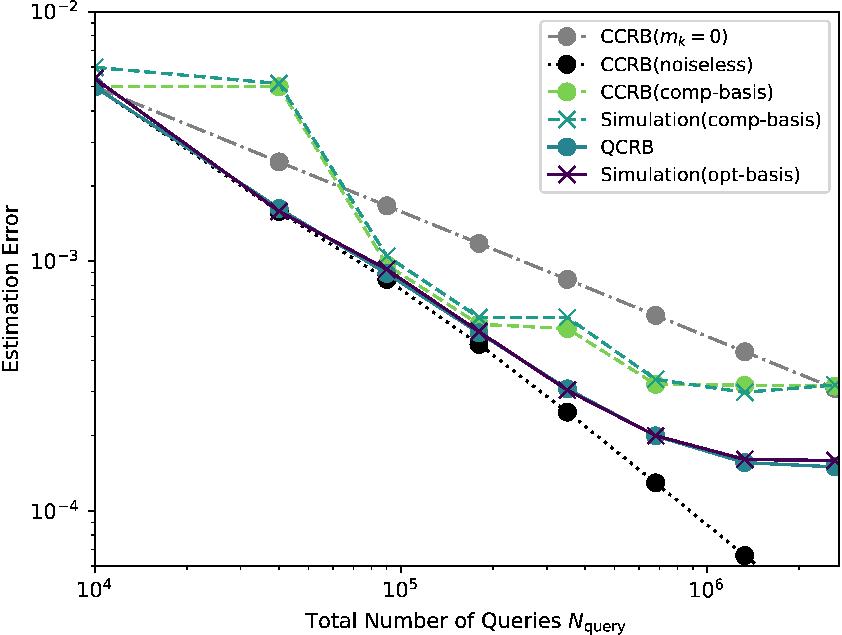}} 
    \subfigure[]{\includegraphics[width=0.99\columnwidth]{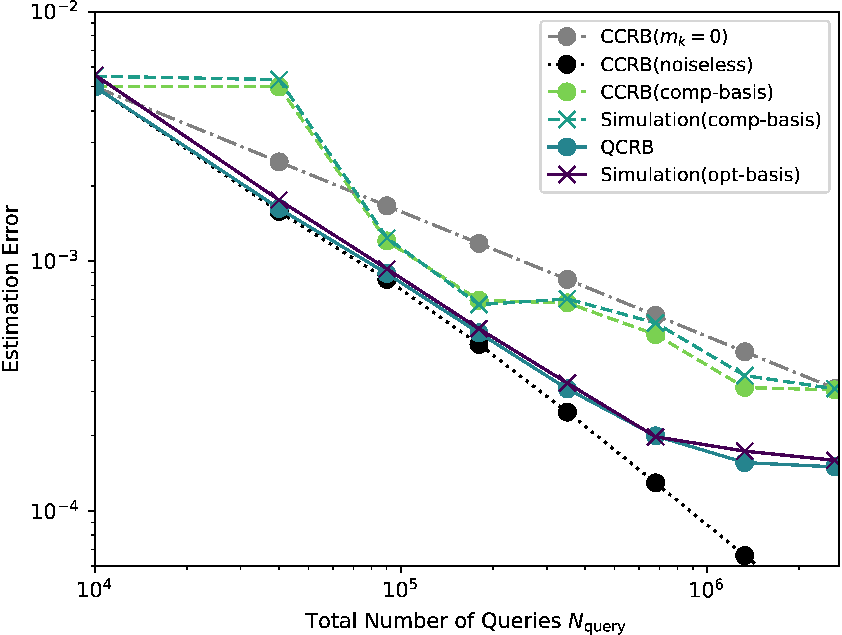}} \hspace{0.02\columnwidth}
    \subfigure[]{\includegraphics[width=0.99\columnwidth]{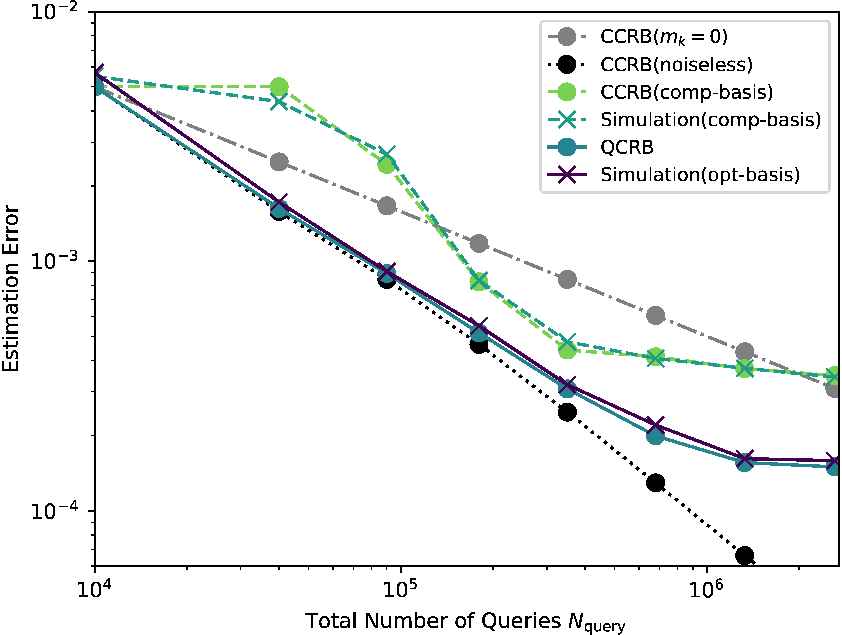}}
\caption{
        \label{Fig_bound_opt-basis}
        The result of the estimation error of $\theta$ with respect to the total number of queries $N_{\rm query} = N_{\rm shot}\cdot N_A$.
        The results for (a), (b), (c), and (d) correspond to the cases where $b_{\rm max}$ = 1/6, 1/4, 1/3, and 2/3, respectively.
        The graph annotated with circle symbols represents the classical Cramér–Rao bounds (CCRB) and quantum Cramér–Rao bounds (QCRB) for estimation error. The cross symbols in green and purple represent the simulation results with computational basis(comp-basis) and optimized basis(opt-basis) measurements. The circle symbols in gray and black represent CCRB for $m_k=0$ (no Grover iteration) and CCRB for $m_k =\{0, 2^0, 2^1, ..., 2^6\}$, each in the absence of noise. The circle symbols in light green and blue show CCRB and QCRB, each in the presence of noise and for $m_k =\{0, 2^0, 2^1, ..., 2^6\}$.
        }
\end{figure*}
We here evaluate the estimation error of $\theta$ via our estimation method. 
The parameter settings for numerical calculations are presented in Table \ref{TABLE_param_adaptive-estimation}.
The number of Grover operators $m_k$ is increased according to the EIS.
The operator $A_{\rm sin}$ in Eq.~(\ref{EQ_operator_A}) takes $b_{\rm max}=1/6, 1/4, 1/3, 2/3$. 
The estimation errors are evaluated by repeating the proposed estimation method $N_{\rm sample}=200$ times for each $m_k$.
The variational parameters ${\bm \alpha}_0$ and ${\bm \alpha}_1$ are optimized in advance, and the nearly optimal operator $C_0({\bm \alpha}_0)$ and $C_1({\bm \alpha}_1)$ (i.e. $|p_{c_0}|$ and $|p_{c_1}| \ge 0.999$) are employed.
We discuss the case when the operation of $C_0({\bm \alpha}_0)$ and $C_1({\bm \alpha}_1)$ are imperfect in Appendix \ref{SEC_estimate_bias}.

\begin{table}[hbtp]
  \caption{List of parameters for the numerical simulation of Section~\ref{SEC_Result_adaptive-estimation}}
  \label{TABLE_param_adaptive-estimation}
  \centering
  \begin{ruledtabular}
  \begin{tabular}{cll}
    number of measurements  & $N_{\rm shot}$  & 10000 \\
    number of qubits  & $n$  & 3 \\
    number of Grover iterations  & $m_k$   & $\{0,2^0,2^1, ... ,2^6\}$ \\
    number of samples  & $N_{\rm sample}$  & 200 \\
    noise parameter & $p$ & 0.95
  \end{tabular}
  \end{ruledtabular}
\end{table}

Figure~\ref{Fig_bound_opt-basis} shows the result of the estimation error of $\theta$ versus the total number of queries $N_{\rm query} = N_{\rm shot}\cdot N_A$. 
From the figures, we find that the estimation error with the conventional method (green cross) is lower bounded by the classical Cramér–Rao bounds (CCRB) (light green circles); on the other hand, the estimation error with our method (purple crosses) nearly attains the QCRB (blue circles), for each $b_{\rm max}$.
It should be noted that the estimation errors at around $N_{\rm query}=10^4$ slightly deviate from QCRBs. 
This discrepancy is caused by the coarseness of the mesh discretization when evaluating the likelihood function $f_L^{\mathrm{MLAE}}(\theta, p; {\bf h})$.

\section{Conclusion}
\label{SEC_Conclusion}
This paper proposes an improved amplitude estimation method to attain the nearly optimal error bound by optimizing the measurement basis.
We assume that the system is affected by depolarizing noise, and the noise intensity is unknown.
Our methodology aims to estimate the parameter $\theta$ embedded in the amplitude and the parameter $p$ representing the noise intensity.
In the amplitude estimation, the optimal measurement basis depends on the unknown quantum states $\ket{\psi_0}_n$, $ \ket{\psi_1}_n$, and the parameter of interest $\theta$.
To address this problem, we employed the adaptive measurement strategy proposed in the context of the quantum estimation theory and adjusted the measurement basis with VQC. 
Notably, since the optimization of the VQC can be performed simultaneously with the first step of amplitude estimation, when this optimization is sufficiently successful, our algorithm can achieve higher accuracy than conventional MLAE with the same number of queries. 
This also means, when the noise intensity is sufficiently small, our algorithm has a query complexity advantage over the classical algorithm, similar to the conventional amplitude estimation algorithms.
We theoretically analyzed the behavior of the optimal bound and found that it is the same as in the case of the estimation of $\theta$ alone.
Additionally, we numerically demonstrated the proposed method using a simulator.
The results show that the estimation error is close to the optimal bound when the VQC works well.
We also observed that, in the proposed method, the optimization of VQC does not suffer from the barren plateau issue.
As mentioned in Section~\ref{SEC_optimize_vqc}, such favorable property in the optimization presumably comes from the locality of the cost function $f_{\rm cost}$ and the shallowness of the VQC.
Furthermore, we examined how much the incomplete optimization of the VQC affects the estimation of $\theta$ through theoretical analysis and numerical calculations.
When the optimization is imperfect, it is necessary to estimate the additional parameters, which represent the ``imperfectness" of the optimization, concurrently with $\theta$ and $p$.
According to the result, even in such a case, the estimation error remains close to the optimal bound when the imperfectness is not large.

Lastly we provide two remarks. 
First, 
in MLAE, it has been known that there are ``anomalous target $\theta$" in which the amplitude amplification with the operator $G$ does not improve the estimation accuracy~\cite{Tanaka2021, Callison2022}.
Since the optimal error bound in Eq.~(\ref{EQ_QCRB}) is independent of $\theta$, the estimation strategy that attains this bound can avoid this problem.
However, in the proposed method, the first-step estimation is performed with the computational basis, thus the estimation accuracy of the anomalous target is not enhanced even with the amplitude amplification.
One way to address this problem is to obtain the estimate of the anomalous target without $G$ and then adjust the measurement basis with this estimate.
Presenting the specific procedures of such approach remains a challenge for future work.

Second, 
recall that we employ the VQC with a specific structure explained in Section~\ref{SEC_optimize_vqc} for adjusting the measurement basis.
Moreover, we can employ an ansatz that takes into account the structure of the operator $A$, which may enable more efficient optimization with less number of parameters. 
Note that such structured VQC could be classically simulable~\cite{Cerezo2023}.
However, in our algorithm, we employ a VQC only for the purpose of adjusting the measurement basis after operating the main quantum circuit; thus, even if the VQC part is classically simulable, this does not mean that the entire quantum circuit can also be classically simulable.

\begin{acknowledgments}
This work is supported by MEXT Quantum Leap Flagship Program Grant Number JPMXS0118067285 and JP- MXS0120319794.
K.W. was supported by JSPS KAKENHI Grant Number JP24KJ1963 and JST SPRING, Grant Number JPMJSP2123.
K.O., N.Y., and S.U. acknowledge support from Center of Innovations for Sustainable Quantum AI from JST, Grant Number JPMJPF2221.
K.O., Y.S., and S.U. acknowledge support from Council for Science, Technology and Innovation(CSTI), Cross-ministerial Strategic Innovation Promotion Program (SIP), the 3rd period of SIP “Promoting the application of advanced quantum technology platforms to social issues”, Grant Number JPJ012367 (Funding agency:QST).
We thank for helpful discussions with Hiroyuki Tezuka, Jumpei Kato, Ryo Nagai, Ruho Kondo, Tatsuhiro Ichimura, Yuki Sato.
We are also grateful to Tamiya Onodera, Tomoki Tanaka for helpful supports and discussions on the early stage of this study.
\end{acknowledgments}

\appendix
\renewcommand{\thetable}{\Alph{section}.\arabic{figure}}
\renewcommand{\thefigure}{\Alph{section}.\arabic{table}}

\section{Detailed derivation of SLD operators for QFIM}
\label{SEC_derivation}
We derive the QFIM for the noisy MLAE discussed in Section~\ref{SEC_Noisy_MLAE}.
First, we obtain the SLD operator $L_\theta$ that satisfies Eq.~(\ref{EQ_def_SLD}) for $\rho_{m_k}(\theta,p)$.
SLD operator is expressed as follows~\cite{Paris2009}:
\begin{align}
    \label{EQ_def_SLD_derivation}
    L_\theta = \sum_j \cfrac{\partial_\theta \lambda_j}{\lambda_j}\ket{\lambda_j}\bra{\lambda_j} + 2\sum_{k \ne l}\cfrac{\lambda_k-\lambda_l}{\lambda_k+\lambda_l} \braket{\lambda_l | \partial_\theta \lambda_k} \ket{\lambda_l}\bra{\lambda_k},
\end{align}
where $\lambda_i , i=0,1,2\cdots2^{n+1}-1$ represent the eigenvalues of $\rho_{m_k}(\theta,p)$, and $\ket{\lambda_i}$ is the corresponding eigenvector.

Here, $\rho_{m_k}(\theta,p)$ is
\begin{widetext}
\begin{align}
    \label{EQ_rho_matrix}
    &\rho_{m_k}(\theta,p) = p^{m_k} \ket{\psi(N_q\theta)}\bra{\psi(N_q\theta)} + (1 - p^{m_k})\cfrac{I_{n+1}}{d} \nonumber \\
    & = \begin{bmatrix} 
          p^{m_k}\cos^2{N_q\theta}+\frac{1-p^{m_k}}{d} & p^{m_k}\cos{N_q\theta}\sin{N_q\theta} & 0 &\dots  & 0 \\
          p^{m_k}\cos{N_q\theta}\sin{N_q\theta} & p^{m_k}\sin^2{N_q\theta}+\frac{1-p^{m_k}}{d} & 0 & \dots  & 0 \\
          0 & 0 & \frac{1-p^{m_k}}{d} & \vdots & 0 \\
          \vdots & \vdots & \vdots & \ddots & \vdots \\
          0 & 0 & 0 & \dots  & \frac{1-p^{m_k}}{d}
        \end{bmatrix}.
\end{align}
\end{widetext}

The eigenvalue $\lambda$ and the eigenvector $\ket{\lambda}$ of $\rho_{m_k}(\theta,p)$ are given by 
\begin{align}
    \label{EQ_eigenvalue}
    \lambda = \left\{\begin{array}{ll}
                        \lambda_0 &= p^{m_k} + \frac{1-p^{m_k}}{d}, \\
                        \lambda_l &= \frac{1-p^{m_k}}{d}  \  (1 \le l \le 2^{n+1}-1),
                     \end{array}
                        \right.
\end{align}
\begin{align}
    \label{EQ_eigenvector}
    \ket{\lambda} = \left\{\begin{array}{lll}
                        \ket{\lambda_0} &= [\cos{N_q\theta}, \sin{N_q\theta},0,...,0]^T, \\
                        \ket{\lambda_1} &= [\sin{N_q\theta}, -\cos{N_q\theta},0,...,0]^T, \\
                        \ket{\lambda_l} &= [0, 0, ..., 1, ..., 0]^T   \  (2\le l \le 2^{n+1}-1). 
                     \end{array}
                        \right.
\end{align}
From Eq.~(\ref{EQ_def_SLD_derivation}), Eq.~(\ref{EQ_eigenvalue}), and Eq.~(\ref{EQ_eigenvector}), $L_\theta$ and $L_p$ is obtained as shown in Eqs.~(\ref{EQ_SLD_theta}, \ref{EQ_SLD_p}).


With $L_\theta$ and $L_p$, we derive $\left[F_q\right]_{\theta_i,\theta_j} (\theta_i,\theta_j=\theta,p)$ and obtain $F_q$ represented in Eq.~(\ref{EQ_QFIM}).

\section{Estimation of the relative phase $\varphi$}
\label{SEC_estimate_varphi}
\begin{figure}[H]
    \centering
    \subfigure[]{\includegraphics[width=\columnwidth]{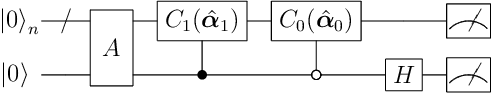}}
    \subfigure[]{\includegraphics[width=\columnwidth]{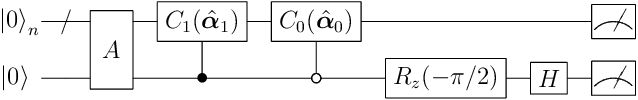}}
\caption{\label{Fig_circuit_estimate_phi} The quantum circuits for estimating $\varphi$ with (a)X-basis measurement and (b)Y-basis measurement.}
\end{figure}

We explain the procedure to estimate $\varphi$ mentioned in Section~\ref{SEC_optimize_vqc}.
As discussed in Section~\ref{SEC_optimize_vqc}, the operator $B'$ is approximately constructed by the VQC.
With this approximation, $B'(\hat{\bm \alpha}_0, \hat{\bm \alpha}_1)$ can yield the operations represented in Eq.~(\ref{EQ_B'_phase}) that result in relative phases $\varphi$, rather than the optimal operation represented in Eq.~(\ref{EQ_operator_B'}).
In this case, for estimating $\varphi$, we measure the ancilla qubit of the quantum state in Eq.~(\ref{EQ_B'_phase}) with $X$ and $Y$-basis, as illustrated in Fig. \ref{Fig_circuit_estimate_phi}, and estimate $\varphi$ with MLE.

Through the measurements, the results are obtained in accordance with the following probabilities:
\begin{align}
    \label{EQ_varphi_estimate_prob}
    {\rm Pr}^X(0;\varphi) &= \cfrac{1}{2} + \cfrac{1}{2}\sin{2\theta}\cos{\varphi}\\
    {\rm Pr}^X(1;\varphi) &= \cfrac{1}{2} - \cfrac{1}{2}\sin{2\theta}\cos{\varphi}\\
    {\rm Pr}^Y(0;\varphi) &= \cfrac{1}{2} + \cfrac{1}{2}\sin{2\theta}\sin{\varphi}\\
    {\rm Pr}^Y(1;\varphi) &= \cfrac{1}{2} - \cfrac{1}{2}\sin{2\theta}\sin{\varphi},
\end{align}
where the superscripts $X$ and $Y$ denote the measurement basis is $X$-basis or $Y$-basis, and the subscripts $0$ and $1$ denote that the measurement result is $0$ or $1$. 
We define the counts of the corresponding results as $h^X_{0,1}$ and $ h^Y_{0,1}$.
With those results, we obtain the estimate $\hat{\varphi}$ which maximizes the likelihood function $f^{\rm phase}_{L}$ such that
\begin{align}
    \label{EQ_varphi_estimate_ll}
    f^{\rm phase}_{L}(h^X_{0,1}, h^Y_{0,1}; \varphi) = \prod_{\sigma \in \{X, Y\}, i\in \{0,1\}} ({\rm Pr}^{\sigma}(i;\varphi))^{h^\sigma_i},
\end{align}
where $0 \leq \varphi \leq 2\pi$.

Below, we analyze the mean square error of $\hat{\varphi}$.
When evaluating $f^{\rm phase}_{L}$, we use the estimate $\hat{\theta}_1$ obtained from the first-step estimation introduced in Section~\ref{SEC_adaptive}.
The estimation error of $\hat{\theta}_1$, defined as $\tilde{\theta}_1 \equiv \hat{\theta}_1 - \theta$, introduces bias into the estimation of $\varphi$, represented as $\varphi_\epsilon$.
From the representation of ${\rm Pr}^X(0;\varphi)$ in Eq.~\eqref{EQ_varphi_estimate_prob}, $\tilde{\theta}_1$ and $\varphi_\epsilon$ are related by
\begin{align}
    \sin{2\theta}\cos{\varphi} = \sin{2(\theta + \tilde{\theta}_1)}\cos{(\varphi + \varphi_\epsilon)}.
\end{align}
When we assume $\tilde{\theta}_1$ is sufficiently small, the approximation
\begin{align}
    \varphi_\epsilon \approx \cfrac{2}{\tan{\varphi}\tan{2\theta}}\cdot\tilde{\theta}_1,
\end{align}
holds.
Performing a similar analysis on ${\rm Pr}^Y(0;\varphi)$ yields
\begin{align}
    \varphi_\epsilon \approx -\cfrac{2\tan{\varphi}}{\tan{2\theta}}\cdot\tilde{\theta}_1.
\end{align}
Since $\hat{\theta}_1$ is estimated from the results of $\sqrt{N_{\rm shot}}$ shots measurements, $\mathbb{E}[\tilde{\theta}_1^2] = \mathcal{O}(1/\sqrt{N_{\rm shot}})$ holds.
Thus, $\mathbb{E}[\varphi_\epsilon^2] = \mathcal{O}(1/\sqrt{N_{\rm shot}})$.
In addition, based on the central limit theorem, the variance of $\hat{\varphi}$ is $\mathcal{O}(1/N_{\varphi-{\rm shot}})$.
Therefore,
\begin{align}
    \label{EQ_varphi_mse}
    \mathbb{E}[(\hat{\varphi} - \varphi)^2] = \mathcal{O}\left( \cfrac{1}{N_{\varphi-{\rm shot}}} + \cfrac{1}{\sqrt{N_{\rm shot}}} \right).
\end{align}

\section{Effect of incomplete $B(\theta)$}
\label{SEC_bias}

We analyze the effect on the estimation of $\theta$ when the construction of $B(\theta)$ is incomplete.
As mentioned in Section~\ref{SEC_optimize_vqc}, the estimation error of $\varphi$ and incomplete optimization of $C_0({\bm \alpha}_0)$ and $C_1({\bm \alpha}_1)$ cause bias and an increase in variance in the estimation of $\theta$.
In such a case, $B(=R_y B')$ performs
\begin{align}
    \label{EQ_varphi_state}
    \ket{\psi} &= \cos{N_q\theta}\ket{\psi_0}_n\ket{0} + \sin{N_q\theta}\ket{\psi_1}_n\ket{1} \nonumber\\
    \xrightarrow{B'} \hspace{4pt} 
          &p_{c_0}\cos{N_q\theta}\ket{0}_n\ket{0} 
           + p_{c_1}e^{i\tilde{\varphi}}\sin{N_q\theta}\ket{0}_n\ket{1} \nonumber\\
          &+ \cos{N_q\theta}\sum^{2^n-1}_{j=1}p_{c_0,j}'\ket{j}_n\ket{0}
           + \sin{N_q\theta}\sum^{2^n-1}_{j=1}p_{c_1,j}'\ket{j}_n\ket{1}  \nonumber\\
         =&p_c\left( \cos{N_q(\theta + \theta_p)}\ket{0}_n\ket{0} 
           + e^{i\tilde{\varphi}}\sin{N_q(\theta + \theta_p)}\ket{0}_n\ket{1} \right) \nonumber\\
          &+ \cos{N_q\theta}\sum^{2^n-1}_{j=1}p_{c_0,j}'\ket{j}_n\ket{0}
           + \sin{N_q\theta}\sum^{2^n-1}_{j=1}p_{c_1,j}'\ket{j}_n\ket{1}  \nonumber\\
    \xrightarrow{R_y} \hspace{4pt} 
          &p_c\left( \cos{(N_q\theta + N_q\theta_p - N_q\hat{\theta}_1-\frac{\pi}{4})}\right. \nonumber\\
          &\left. \hspace{5pt} - \sin{(N_q\theta+N_q\theta_p)}\sin{(N_q\hat{\theta}_1+\frac{\pi}{4})}(1-e^{i\tilde{\varphi}}) \right) \ket{0}_n\ket{0}\nonumber\\
         +&p_c\left( \sin{(N_q\theta + N_q\theta_p - N_q\hat{\theta}_1 - \frac{\pi}{4})}\right. \nonumber\\
          &\left. \hspace{5pt} - \sin{(N_q\theta+N_q\theta_p)}\cos{(N_q\hat{\theta}_1+\frac{\pi}{4})}(1-e^{i\tilde{\varphi}}) \right) \ket{0}_n\ket{1}\nonumber\\
          &+ \cos{N_q\theta}\sum^{2^n-1}_{j=1}p_{c_0,j}'\ket{j}_n R_y \ket{0} \nonumber \\
          &+ \sin{N_q\theta}\sum^{2^n-1}_{j=1}p_{c_1,j}'\ket{j}_n R_y \ket{1},
\end{align}
where $R_y$ denotes $R_y(-2N_q\hat{\theta}_1-\frac{\pi}{2})$. 
$p_c$ is defined as $|p_c|^2 \equiv |p_{c_0}|^2\cos^2{N_q\theta} + |p_{c_1}|^2\sin^2{N_q\theta}$, and means how much the operation of $C_0(\hat{\bm \alpha}_0)$ and $C_1(\hat{\bm \alpha}_1)$ are inaccurate.
While $p_c$ is complex, the phase does not affect the analysis; therefore, only the absolute value is defined here.
$\tilde{\varphi}$ denotes the estimation error of $\varphi$(i.e. $\tilde{\varphi} \equiv \varphi - \hat{\varphi}$ ), and $\theta_p$ denotes the bias caused by $C_0(\hat{\bm \alpha}_0)$ and $C_1(\hat{\bm \alpha}_1)$.

First, we consider the bias $\theta_p$ described by 
\begin{align}
    \label{EQ_bias_p}
    \theta_p &= \cfrac{1}{N_q}\arctan{\cfrac{\left( \cfrac{p_{c_1}}{p_{c_0}} - 1 \right) \tan{N_q\theta}}{1 + \cfrac{p_{c_1}}{p_{c_0}}\tan^2{N_q\theta}}} \nonumber \\
    &\approx \cfrac{1}{2N_q}\sin{2N_q\theta}\left( \cfrac{p_{c_1}}{p_{c_0}}-1 \right).
\end{align}
If there is a bias in the optimization of $C_0({\bm \alpha}_0)$ and $C_1({\bm \alpha}_1)$ (i.e. $p_{c_0}\ne p_{c_1}$), then $\theta_p \ne 0$ holds.
To mitigate the effects of this bias, it is necessary to estimate $p_{c_0}$ and $p_{c_1}$.
We provide the details about this estimation method in Appendix~\ref{SEC_estimate_bias}.

Next, we examine the variance in the estimation of $\theta$.
When we measure the quantum state $\ket{\psi}$ in Eq.~(\ref{EQ_varphi_state}) with the measurement bases $\ket{\lambda_0(\theta)}_{n+1},\ket{\lambda_1(\theta)}_{n+1}, \{\ket{\lambda_l}_{n+1}\}$, the probability of the measurement results are 
\begin{align}
    \label{EQ_probability_optbasis_error}
    {\rm Pr}(\lambda_0 ; \theta, p, m_k) \approx \; & p^{m_k}|p_c|^2 \frac{1}{2} \left[ 1 + \sin{2N_q \left(\theta-\hat{\theta}_1+\theta_p\right)}  \right. \nonumber \\
    &\left. - \frac{1}{2} \tilde{\varphi}^2 \sin{2N_q\left(\theta +\theta_p \right)} \cos{2N_q\hat{\theta}_1}  \right] \nonumber \\
    & +\frac{1-p^{m_k}}{d} \nonumber \\
    {\rm Pr}(\lambda_1 ; \theta, p, m_k) \approx \; & p^{m_k}|p_c|^2 \frac{1}{2} \left[ 1 - \sin{2N_q \left(\theta-\hat{\theta}_1+\theta_p\right)} \right. \nonumber \\
    &\left. + \frac{1}{2} \tilde{\varphi}^2 \sin{2N_q\left(\theta +\theta_p \right)} \cos{2N_q\hat{\theta}_1}  \right] \nonumber \\
    & +\frac{1-p^{m_k}}{d} \nonumber \\
    {\rm Pr}(\lambda_l ; \theta, p, m_k) = \; & \cfrac{(1-p^{m_k})}{d}(d-2) + p^{m_k}(1 - |p_c|^2). 
\end{align}
Here, we assume $\tilde{\varphi}$ is sufficiently small and approximate as $\cos{\tilde{\varphi}} \approx 1-\frac{1}{2}\tilde{\varphi}^2$.
As these equations indicate, $|p_c|^2$ has the same effect on ${\rm Pr}(\lambda_0)$ and ${\rm Pr}(\lambda_1)$ as the noise intensity $p$ and increase the variance in the estimation of $\theta$.
Nevertheless, as mentioned in Section~\ref{SEC_optimize_vqc}, the effect of $|p_c|^2$ is independent of the number of Grover operations $m_k$, while the effect of $p$ increases exponentially depending on $m_k$.
For this reason, $|p_c|^2$ is negligible when $m_k$ is large.

Here, we assess the effect of the estimation error $\tilde{\varphi}$.
Based on Eq.~(\ref{EQ_probability_optbasis_error}), $\tilde{\varphi}$ causes the bias $\theta_\varphi$ represented by
\begin{align}
    \label{EQ_varphi_bias1}
    &\sin{2N_q \left(\theta-\hat{\theta}_1+\theta_p\right)} - \frac{1}{2} \tilde{\varphi}^2 \sin{2N_q\left(\theta +\theta_p \right)} \cos{2N_q\hat{\theta}_1} \nonumber \\
    & = \sin N_q \left[ 2\left( \theta-\hat{\theta}_1+\theta_p \right) + \theta_\varphi \right]. 
\end{align}
Then, $\theta_\varphi$ can be approximated as
\begin{align}
    \label{EQ_varphi_bias2}
    \theta_\varphi \approx -\cfrac{1}{N_q \sqrt{1-\sin^2{2N_q(\theta-\hat{\theta}_1 + \theta_p)}}} \cdot \gamma,
\end{align}
where $\gamma=\frac{1}{2} \tilde{\varphi}^2 \sin{2N_q\left(\theta +\theta_p \right)} \cos{2N_q\hat{\theta}_1}$, and there is a linear approximation for  $\gamma$. 
Since $\theta - \hat{\theta}_1 \approx 0$ and $\theta_p\approx0$, the approximation $\sqrt{1-\sin^2{2N_q(\theta-\hat{\theta}_1+\theta_p)}} \approx 1$ holds.
Based on Eq.~\eqref{EQ_varphi_mse}, $\gamma = \mathcal{O}(\tilde{\varphi}^2)=\mathcal{O}(1/N_{\varphi-{\rm shot}} + 1/\sqrt{N_{\rm shot}})$.
Therefore, 
\begin{align}
    \theta_\varphi = \cfrac{1}{N_q} \cdot \mathcal{O}\left( \cfrac{1}{N_{\varphi-{\rm shot}}} + \cfrac{1}{\sqrt{N_{\rm shot}}}\right).
\end{align}
For estimation of $\theta$, the lower bound of the estimation error is $\mathcal{O}(1/N_q \sqrt{N_{\rm shot}})$.
Thus, when we take the $N_{\varphi-{\rm shot}}$ as $N_{\varphi-{\rm shot}}=\mathcal{O}(\sqrt{N_{\rm shot}})$, the effect of $\tilde{\varphi}$ can be sufficiently reduced.
This means that when $N_{\rm shot}$ is sufficiently large, we can mitigate $\theta_\varphi$ with negligible overhead compared to $N_{\rm shot}$.

\section{Estimation of the bias $\theta_p$ and the change of QCRB}
\label{SEC_estimate_bias}
In this appendix, we show the method to reduce the effect of the bias $\theta_p$ by estimating $p_{c_0}$ and $p_{c_1}$.

Here, we construct the likelihood function $f_L(\theta, p, p_{c_0}, p_{c_1}; {\bf h})$ same as Eq.~(\ref{EQ_loglikelihood}) with the measurement probabilities in Eq.~(\ref{EQ_probability_optbasis_error}).
With this likelihood function, we can estimate four parameters $\theta, p, p_{c_0}, p_{c_1}$ simultaneously by the same method introduced in Section~\ref{SEC_optimal_measurement_basis}.

We demonstrated this estimation numerically and examined the shift of the lower bound of the estimation error and the attainability of that lower bound.
The parameters for the numerical simulation are the same as Table~\ref{TABLE_param_adaptive-estimation}, while $C_0({\bm \alpha}_0)$ and $C_1({\bm \alpha}_1)$ are incomplete ($|p_{c_0}| \approx 0.942, |p_{c_1}| \approx 0.880$).

\begin{figure}[H]
    \centering
    \includegraphics[width=\columnwidth]{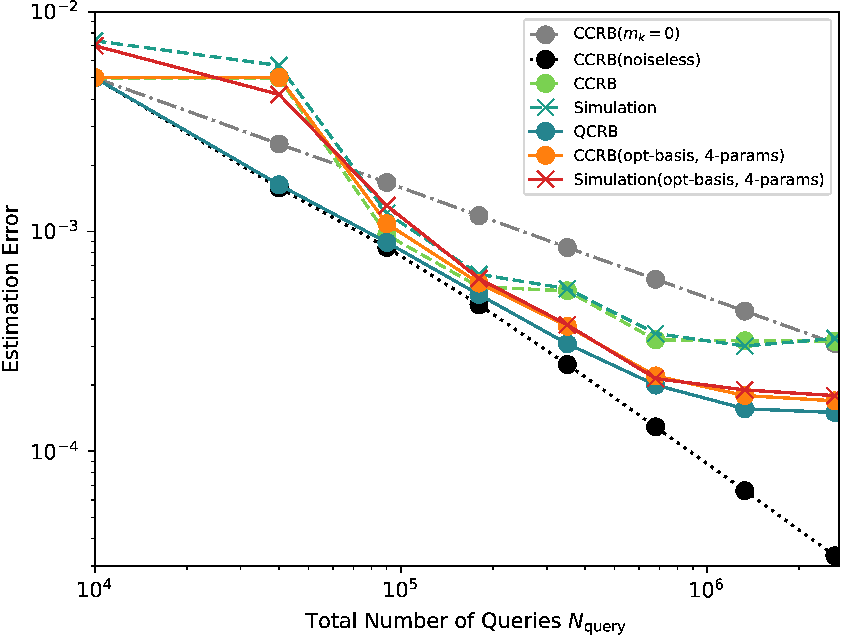}
    \caption{\label{Fig_bound_opt-basis_4params} The result of the estimation error of $\theta$ with respect to the total number of query $N_{\rm query} = N_{\rm shot}\cdot N_A$ for the simultaneous estimation of $\theta, p, p_{c_0}, p_{c_1}$. The circle symbol in orange shows CCRB for the simultaneous estimation of $\theta, p, p_{c_0}, p_{c_1}$. The cross symbol shows the simulation result. Other symbols follow the same representation as Fig. \ref{Fig_bound_opt-basis}.
    }
\end{figure}

The result shown in Fig.~\ref{Fig_bound_opt-basis_4params} represents that, even when we estimate $\theta, p, p_{c_0}, p_{c_1}$ simultaneously, the lower bound remains close to the original lower bound if $p_{c_{0,1}}$ is around $0.9$. 
In addition, the result shows that this shifted lower bound is attainable in such a case.

\bibliography{main}

\end{document}